\documentclass[aps,prb,twocolumn]{revtex4}
\usepackage{amssymb}

\usepackage{graphicx}% Include figure files

\newcommand{\imag}[0]{\dot{\imath}}

\begin{document}
 
\title{Enhancement of Wigner crystallization in quasi low-dimensional solids.}

\author{G. Rastelli$^{1}$} \author{P. Qu\'emerais$^2$} \author{S. Fratini$^2$}

\affiliation{$^{1}$ Istituto Nazionale di Fisica della Materia and
Dipartimento di Fisica\\
Universit\`a dell'Aquila,
via Vetoio, I-67010 Coppito-L'Aquila, Italy}

\affiliation{$^{2}$  Laboratoire d'Etudes des
Propri\'{e}t\'{e}s Electroniques des Solides,
CNRS \\ BP 166 - 25, Avenue des Martyrs, F-38042 Grenoble Cedex 9, France}

\begin{abstract} 
The crystallization of electrons in quasi low-dimensional
solids is studied in a model which retains the full three-dimensional nature
of the Coulomb interactions.
We show that restricting the electron motion to layers (or chains) 
gives rise to  a rich sequence of structural transitions 
upon varying the particle density.
In addition, the concurrence of low-dimensional electron motion and 
isotropic Coulomb interactions leads to a sizeable stabilization 
of the Wigner crystal, which 
could be one of the mechanisms at the origin of
the charge ordered phases frequently  observed in such compounds.
\end{abstract}

\date{January 25, 2006}
\maketitle

\section{Introduction}

Despite being a well established  concept in the physics of
interacting electrons, direct evidence of Wigner crystallization \cite{Wigner} 
has been reported unambiguously in only a limited number of systems, 
namely 
electrons at the surface of liquid
helium, \cite{Crandall_Williams,Grimes_Adams,Andrei_book} and 
in semiconductor heterostructures
of extreme purity. \cite{Yoon}
In both cases, 
a  two dimensional electron gas (2DEG)  
is realized at an interface between two media, 
for which the jellium model of the homogeneous electron 
gas
constitutes a good approximation.

Alternatively, 
the charge ordering phenomena 
observed at low temperatures in a number of \textit{solids}
have been either
interpreted as some form of Wigner crystallization, or
ascribed to the presence of long-ranged Coulomb interactions. 
These include the one-dimensional organic salts
TTF-TCNQ, \cite{Hubbard78} 
(TMTTF)$_2$X, \cite{Monceau,Chow} (DI-DCNQI)$_2$Ag, \cite{Hiraki-Nakazawa}
the ladder cuprate compounds Sr$_{14}$Cu$_{24}$O$_{41}$
\cite{Abbamonte,Abbamonte2} and chain compounds Na$_{1+x}$CuO$_2$,\cite{Horsch}
as well as the  layered superconducting  cuprates 
\cite{cuprates,holecrystal}   and possibly the two-dimensional
BEDT-TTF  organic salts.  \cite{Drichko}  For such systems,  
the jellium model is \textit{a  priori} a rather crude modelization, 
and the concept of Wigner crystallization must be generalized to account for 
other competing effects such as
the periodic potential of the underlying lattice, chemical
impurities, structural defects, magnetic interactions, etc. 
In narrow band solids, for instance, 
the interplay with the host lattice of ions
can strongly affect the charge
ordering pattern especially at highly commensurate band
fillings. \cite{Hubbard78,Baer05}
Nevertheless,
when  the radius of localization of the particles  is larger than the 
typical ion-ion distance, 
the host lattice can be replaced  to a good
accuracy  by an effective continuous
medium, restoring \textit{de facto} the validity of the jellium model.
\cite{Pichard,Valenzuela}

Setting aside  
the important problem of the commensurability with the host lattice, 
and neglecting  disorder and other effects that can certainly play a
role in the compounds under study, we come to the following observation:
a common feature shared by the experimental systems listed above  
is that they are all \textit{quasi low-dimensional} solids, i.e.
they are bulk three-dimensional (3D)
compounds  where the transfer integrals between different
chemical units are so anisotropic that the
carrier motion is effectively restricted to two-dimensional
(2D) atomic layers, or one-dimensional (1D) chains.
Yet, the Coulomb forces 
retain their  three-dimensional character, being 
long-ranged and
\textit{isotropic}. In such systems,  inter-layer 
(inter-chain) interactions cannot be
neglected, leading  eventually to a full three-dimensional 
ordering of the charges. \cite{Abbamonte2,holecrystal}
This suggests
why quasi low-dimensional solids are a 
particularly favorable ground for the
observation of Wigner crystallization:
the electron-electron interactions have the same behavior as in
bulk three-dimensional systems, but the kinetic part is strongly reduced by the
effective lowering of dimensionality. Reminding that a Wigner
crystal arises from the competition  between potential and
kinetic energy, this results in a sizeable
stabilization of the crystal  as compared with  the usual 3D case.
\footnote{The kinetic energy evaluated in the noninteracting limit 
is strongly reduced in lower dimensions, where there are less high
momentum states available
($T=1.11/r_s^2$,  $0.5/r_{s,2D}^2$ and $0.11/r_{s,1D}^2$ in 3D, 2D and 1D respectively)
while the Madelung energy in the opposite classical limit
is less dependent on dimensionality
[$E_{3D}=0.896/r_s$ for the body centered cubic lattice,
$E_{2D}=1.11/r_{s,2D}$ for the hexagonal lattice]. 
Incidentally, the crystallization transition in both 3D ($r_s\simeq 100$) and 
2D ($r_{s,2D}\simeq 40$) takes place at the same value of the ratio
$E/T\simeq 81$.}

A similar conclusion is reached by observing that, even compared to
purely low-dimensional systems such as the 2DEG mentioned above,  
the  Wigner crystal phase  could be 
stabilized  in quasi low-dimensional solids
 due to the presence of additional interlayer interactions.
This 
topic has been analyzed in the literature in the  
framework of bilayer quantum wells, i.e.  constituted of two 
coupled 2D electron systems, where 
it has been shown that, depending on the 
strength of the interlayer forces,
the ordering pattern 
can differ
from the hexagonal structure expected in a single layer.
\cite{Goldoni_Peeters,Schweigert_Schweigert_Peeters} 
More importantly, it was
found\cite{swierkowski,contiIJMPB,Goldoni_Peeters_2}
that at  interlayer
separations comparable with the mean interparticle distance, 
the melting density is raised by a factor of 3 
with respect to the pure 2D case, which makes a factor as large as $10^2$ when
appropriately scaled  to the 3D situation. 

In this work, we model quasi two-dimensional (one-dimensional) systems as
periodic arrays of conducting layers (wires) embedded in a
three-dimensional bulk material, where the electrons
interact through isotropic long-range Coulomb forces.
We show that, upon  varying the particle density or the interlayer (interwire) separation, the Wigner crystal undergoes 
several structural transitions  in order to minimize 
its energy compatibly with the given geometrical constraints.
We then give a semi-quantitative estimate of the melting density 
for the different structures previously 
identified,  based on the 
Lindemann criterion, which confirms  the stabilization 
of the crystallized phase expected from general grounds.

The paper is organized as follows:
In Section II, we introduce a model for the crystallization of
electrons in an
anisotropic environment and the method for calculating the
crystal energy in the harmonic approximation, which includes the 
classical Madelung energy and the zero-point vibrational energy
of the collective excitations. This is applied 
to the case of  quasi two-dimensional  systems, for which the
structural/melting phase diagram is determined. The validity of the
present  
approximation scheme is checked at the end of Section II 
by analyzing a system of two coupled layers, for which  our results 
compare positively  with the numerical results available in the literature.
An analogous discussion for quasi one-dimensional systems is reported in
Section III, by treating explicitely the case where the conducting chains 
form a square array. The main results are summarized in section IV.

\section{Wigner crystallization in layered solids}

\subsection{Model and approximations}

Let us  consider a system of electrons (or holes)  of density $n=(4
\pi r_s^3/3)^{-1}$  
in a strongly anisotropic environment, such  that the particle motion
is  constrained to 
equally spaced atomic layers (at distance $d$), 
but remains isotropic within the layers.
To ensure charge neutrality, we assume a uniform 3D compensating
background of opposite charge. 
The hamiltonian for $N$ crystallized particles in a volume $V$ is given by:
\begin{equation} 
\label{energy}
H = N E_M + \sum^{N}_{i=1} \frac{p^2_{i}}{2 m}  + V_d
\end{equation}
The first term  
\begin{equation}
\label{eqn:E_Mad}
E_M= \frac{e^2}{2}  \left[ \sum_{i} \frac{1}{R_{i}} -  n
  \int_V \!\! \frac{d \vec{r}}{r} \right] \label{eqn:V_o}
\end{equation}
is the Madelung energy of the given lattice structure 
(in the thermodynamic limit, $N,V \rightarrow \infty$, boundary effects
are negligible and all particles become  equivalent). The second term
is the \textit{two-dimensional} 
kinetic energy of the localized particles and the last term 
accounts for the interactions due to 
the \textit{planar} 
displacements  $\vec{u}_{i}=(u_{xi},u_{yi})$ of the
electrons around their equilibrium positions:
\begin{equation}
V_d  =    
\frac{e^2}{2} \sum_{i \neq j} \left[
\frac{1}{\left| \vec{R}_{i} + \vec{u}_{i} - \vec{R}_{j} - \vec{u}_{j} \right|} -
\frac{1}{\left| \vec{R}_{i} - \vec{R}_{j}  \right|}
\right]
\end{equation}

Expanding the last term for small displacements results in a series
expansion  
for the energy Eq. (\ref{energy}) in powers of  $1/r_s^{1/2}$.
\cite{Carr,Carr_Coldwell-Horsfall_Fein} 
The leading term, proportional to $1/r_s$, corresponds to
the Madelung energy $E_M$ of Eq.(\ref{eqn:E_Mad}).
In free space, it attains its minimum value
$E_{BCC}=-0.89593/r_s$ (in atomic units) 
for a Body Centered Cubic (BCC)  Wigner crystal.
\cite{Clark} 
The second term in the expansion, proportional to
$1/r_s^{3/2}$, is the 
zero point energy of the particle fluctuations
in the harmonic approximation, which also depends on the selected crystal
structure. It is negligible for $r_s\to \infty$, and remains 
smaller than the Madelung term by typically an order of
magnitude at  $r_s\sim 100$.
Nonetheless, 
it can play an important role in
determining the relative stability
of the different crystal structures, especially when approaching
the melting density.
Higher orders in the energy expansion \cite{Carr,Carr_Coldwell-Horsfall_Fein}  
include anharmonic ($1/{r_s}^p$ with $p\geq 2$) and  exchange terms of the form
$e^{-c\sqrt{r_s}}$, 
which we shall neglect in the following discussion. 

Up to quadratic order in the displacements, our model Hamiltonian  reads:
\begin{equation} \label{Hquad}
  H=N E_M+\sum_i \frac{p^2_{i}}{2 m} + \frac{e^2}{4} \sum_{i,j\neq i}
\left(  \vec{u}_i - \vec{u}_j \right) \hat{I}_{ij} \left(  \vec{u}_i - \vec{u}_j \right)
\end{equation}
where $\hat{I}_{ij}$ is a  $2 \times 2$ matrix characterizing the
dipole-dipole interactions, given by
($\alpha,\beta=x,y$): 
\begin{equation}
  \label{eqn:I_matrix}
{ \left( \hat{I}_{ij} \right) }_{\alpha \beta}
= 
\frac{3 \vec{R}_{i j ,\alpha} \vec{R}_{i j ,\beta}}{ {\left| {R}_{i j } \right|}^5 }  -
\frac{\delta_{\alpha \beta}}{ {\left| {R}_{i j } \right|}^3 } 
\end{equation}
with $  \vec{R}_{i j } = \vec{R}_{i} - \vec{R}_{ j } $.
The most general elementary Bravais lattice compatible with a given layered
structure is identified by a couple of basis vectors describing the
ordering within the planes, $\vec{A}_1=(a_1,0,0)$, $\vec{A}_2=(a_{2x}
,a_{2y},0)$, and a third vector
$\vec{A}_3=(a_{3x},a_{3y},d)$ which sets the relative shift ($a_{3x},a_{3y}$)
between two equivalent 2D-lattices on neighboring planes.
Other structures, with more than one particle per unit cell, 
are  possible  in principle, but will not be considered here.
 
Due to the additional lengthscale $d$ introduced by the layered
constraint, the crystal energy is no longer a function of $r_s$ alone.
Its dependence on the lattice geometry is best expressed by introducing  
a dimensionless parameter $\gamma$, which measures of the
relative importance of  interlayer and intralayer interactions. It is
defined as the ratio between the mean 
interparticle distance in the planes  and the  interlayer
separation, namely  $\gamma =  \sqrt{\pi} r_{s,2D}/d$.
Here $r_{s,2D}$ defines the 2D  density parameter in
the individual layers, related to  the bulk $r_s$ by $r_{s,2D}^2=4 r_s^3/3d$.
The first two terms of the low-density expansion, corresponding
respectively to the Madelung energy and the zero-point fluctuation
energy in the quadratic model (\ref{Hquad}) can be written in compact
form as:  
\begin{equation}
\label{eq:crysten}
  E=\frac{A(\gamma)}{r_s}+\frac{B(\gamma)}{r_s^{3/2}}.
\end{equation}

It should be noted that  an effective mass $m^*\neq m$
and a dielectric constant $\kappa\neq 1$
can be straightforwardly included in the model through a redefinition of the
Bohr radius $a_B \to a_B^*= a_B  \kappa (m/m^*)$, unit energy
$ m e^4/\hbar^2\to m^*e^4/\kappa^2\hbar^2$, and
density parameter  $r_s\to r_s (m^*/m)/\kappa$.  
Hereafter, energies and lengths will therefore be expressed in terms of these
effective units, characterizing the host medium.
A much more  complex situation arises in systems with a frequency-dependent 
dielectric screening, leading to the formation of polarons, 
for which the reader is referred to 
Refs.\cite{holecrystal,PWC}.

\subsection{Minimization of the Madelung energy}

Following the hierarchy of the 
series expansion introduced above, 
we start by searching  for the layered configuration which minimizes the 
electrostatic repulsion between the particles, which is appropriate 
in the limit of large $r_s$.
The calculation is performed by standard Ewald summation techniques,
which split the slowly convergent series in  Eq. (\ref{eqn:V_o}) into
two exponentially converging sums. \cite{Ewald}
Given the interlayer separation $d$ and the bulk density $n$ (or,
alternatively, given the pair of dimensionless parameters $\gamma$ and $r_s$)
we are left with 4 free minimization  parameters: 
2 for the inplane structure, 2 for the interlayer ordering.

\begin{figure}[htbp]
\centering 
\includegraphics[scale=.35,angle=270.]{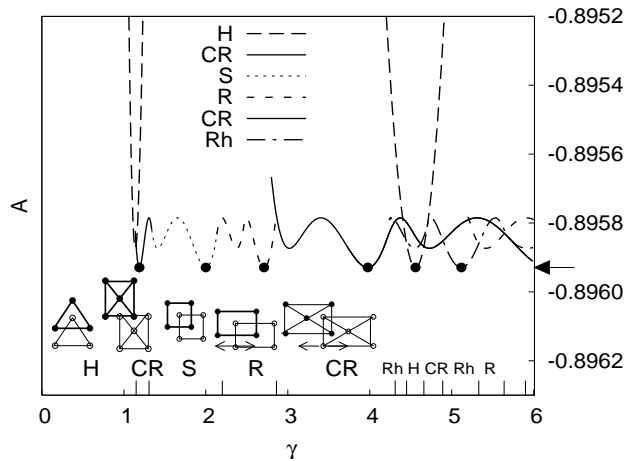}
\caption{Madelung coefficient $A$ in atomic units for different crystal
structures constrained to a layered environment, 
as a function of the anisotropy ratio $\gamma$.  
The different curves correspond to different planar
configurations: 
hexagonal (H), square (S),  
centered rectangular (CR), 
rectangular (R) and  rhombic (Rh). 
The interlayer orderings in the simplest cases 
at low $\gamma$ are sketched below the curves
(for the R and CR structures, the stacking 
varies as indicated by the double arrows). 
The resulting three-dimensional Wigner crystal reduces to a perfect  
BCC at the points marked by filled dots, whose energy is
indicated by the  horizontal  arrow.}
\label{fig:madelung}
\end{figure}

The result of the minimization for the Madelung coefficient $A$ 
in the range $0<\gamma<6$ is illustrated in Fig. \ref{fig:madelung}.
Two distinct regimes can be identified. 
In the limit of  large separations ($\gamma\lesssim 1$), 
the coupling between the layers is weak, and the resulting planar pattern is
hexagonal, with a  staggered interlayer ordering, i.e.  
the particles on the neighboring layers
falling on top of the centers of the triangles.
The sharp rise of the  Madelung constant 
in this regime
is due to the fact that
the  compensating background is distributed
homogeneously in three-dimensional space, which 
penalizes strongly anisotropic charge distributions.
\footnote{  
The usual value for the Madelung energy of the hexagonal lattice 
in two dimensions would be recovered at large interlayer separations
by considering a layered compensating charge. This would not alter 
the sequence of crystal structures, since it would add
a constant term to the energy depending only on the anisotropy ratio $\gamma$.}

Upon reducing the interlayer separation so that $\gamma \gtrsim 1$, 
the increasing interlayer
interactions make the  hexagonal pattern energetically unfavorable. 
Above $\gamma=1.15$, a more isotropic ordering of the charges is
stabilized, which presents a centered rectangular (CR) structure in
the planes.   
Further increasing $\gamma$
leads to a sequence of structures whose planar patterns
are respectively squared (S, 
in the interval $1.32<\gamma<2.13$), rectangular
(R,  $2.13<\gamma<2.84$), 
centered rectangular (CR, $2.86<\gamma<4.31$), 
a generic rhombic, or oblique phase 
(Rh, $4.31<\gamma<4.45$), then hexagonal again, and so on. 
Such phases are  all connected by continuous structural
transitions, 
with the exception of the hexagonal structure, 
which is attained through a discontinuous change of the crystal
parameters.
% (although the energy curve remains continuous).
Note that in the very narrow interval
$2.84<\gamma<2.86$, a generic structure with rhombic planar
symmetry is stabilized, which 
allows to evolve continuously from the  rectangular to the 
centered rectangular patterns (not shown).
The sequence of structural transitions goes on at larger values of
$\gamma$.

The interlayer ordering is shown at the bottom of  Fig.\ref{fig:madelung}.
It is staggered for the first three patterns (H), (CR) and (S)  
for $\gamma\lesssim 2$, 
as expected for large interlayer separations, where 
the relative ordering is fully determined by the coupling
between two adjacent planes, and indeed coincides with what is found
in  bilayer systems \cite{Goldoni_Peeters} (see Section
\ref{sec:bilayer} below). At larger values of $\gamma$, 
the interactions beyond the nearest planes  become
relevant, which makes the simple staggered ordering  unfavorable.  For 
instance, a staggered/non-staggered transition takes place within the 
rectangular phase at $\gamma =2.46$, corresponding to a relative sliding
of the planar structures on adjacent planes in the
direction of the long bonds (indicated by the double arrow in Fig. 1).

Remarkably, each of the phases identified above contains a special point  
$\gamma^*$ where the ideal BCC structure ---which has the lowest
possible Madelung energy in three dimensions---
is itself compatible with the layered constraint. 
The different planar configurations
identified above then correspond to the different ways of
cutting a BCC by an array of equally spaced layers.
Such points are easily calculated by setting the
distance $d=2\pi/|{\bf K}|$, with ${\bf K}$ any reciprocal lattice vector, 
and correspond to  $\gamma^*=2^{1/4}$, $2$,
$2^{1/4}  3^{3/4}$,$2^{1/4}  5^{3/4}$, $2 \
3^{3/4}$, etc\ldots Similarly, 
the higher relative minima visible in Fig. \ref{fig:madelung} correspond
to different orientations of the same 
three-dimensional Face Centered Cubic (FCC) ordering.

Away from such special points, the overall charge distribution  
remains very isotropic in all the region $\gamma\gtrsim 1$, 
as testified by the extremely small deviations of the  Madelung
energy  from the ideal case, 
$\Delta E_M \lesssim  10^{-4}/r_s$.
Such small energy variations, however,  refer to the  optimal
structures obtained at different values of $\gamma$, which
\textit{does not mean} that the electrostatic repulsion between the carriers 
is irrelevant in the
determination of the charge ordering patterns in real systems:
in a given compound, 
where both  the interlayer distance and the density are fixed, one 
should rather 
compare the energies of two competing phases at \textit{fixed}
$\gamma$. For example, enforcing a hexagonal symmetry  at $\gamma=2$,  
where the optimal structure is squared,
would cost an energy 
$\Delta E_M\sim 0.015/r_s \sim 200K$ at $r_s=20$, 
which is comparable with the typical charge ordering energy scales in
solids. Yet, since the Madelung energy 
is determined by the interactions with a large number of (distant)
neighbors, the structures found here are expected to 
be relatively soft against local deformations. The situation is
different regarding global symmetry changes, as can result from
the inclusion of a periodic potential of competing symmetry, which could
strongly modify the sequence and order of the structural transitions,
possibly favoring the appearance of alternative phases.
\cite{Sherman,Tsiper}

\subsection{Zero point fluctuation energy}

\begin{figure}[htbp]
\centering
\includegraphics[scale=.35,angle=270.]{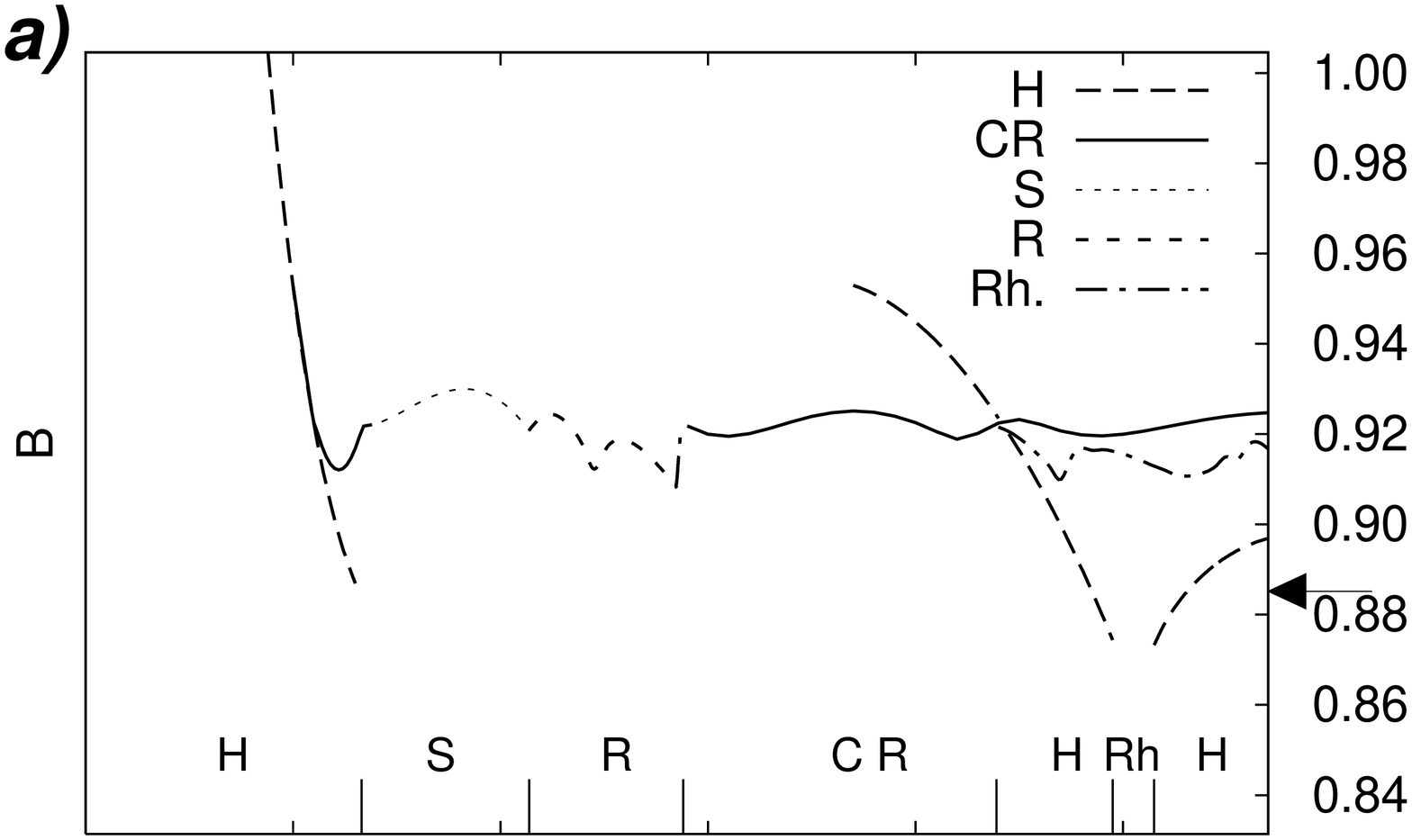}
\includegraphics[scale=.34,angle=270.]{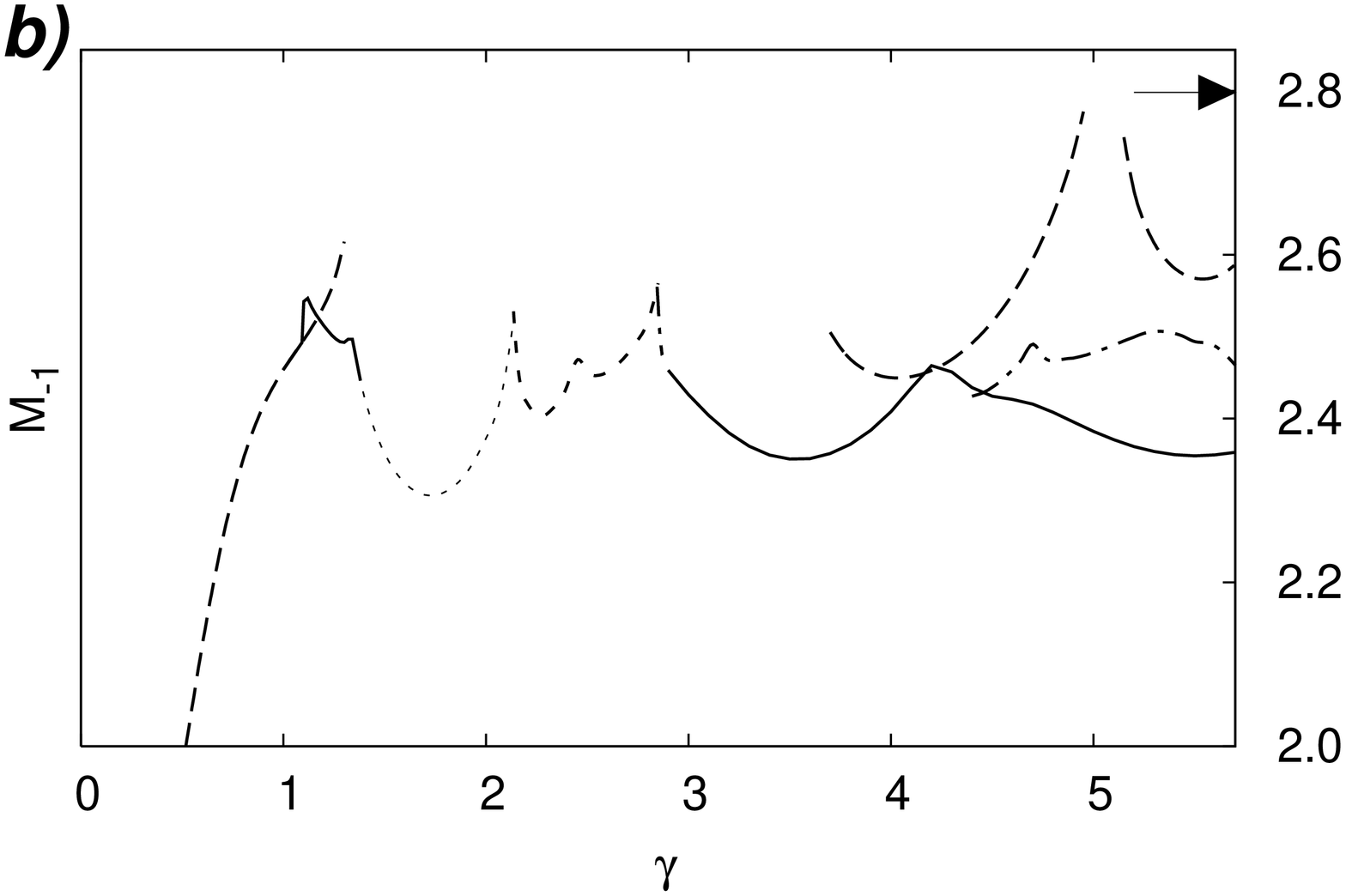}
\caption{a) Zero point vibrational term $B$ 
for the different structures identified in
Fig. 1, within their ranges of mechanical stability. 
The sequence of structures with the lowest vibrational energy is
indicated at the bottom.
The arrow indicates the value  
$(2/3)B^{(3D)}=0.887$, where $B^{(3D)}$ is the vibrational energy of a BCC
crystal in vacuum; 
b) Inverse moment $M_{-1}$ of the DOS, 
which is proportional to the mean electronic fluctuation $\langle u^2\rangle$   
Eq.(\ref{eqn:u2}) for the same structures. 
The arrow indicates the value for a BCC in free space.}
\label{fig:B_coefficient}
\end{figure}

The next term in the series expansion of the ground state energy 
Eq. (\ref{eq:crysten}) corresponds to the quantum zero point
fluctuations of the particles around their equilibrium positions,
in the harmonic approximation.  It is negligible at large $r_s$ (low
density),  but it
becomes quantitatively important at lower $r_s$, where it can 
slightly modify
the sequence of phases identified in the preceding Section.  
Upon further reducing $r_s$, this term eventually drives 
the quantum melting of the crystal, that will be analyzed in the next
Section.

The calculation of the fluctuation term proceeds as follows.
The harmonic model  Eq.(\ref{Hquad}) is diagonalized by 
introducing the normal modes $q_{s,\vec{k}}$ 
\begin{equation} \label{eqn:q_ks}
\vec{u}_{i}  = \frac{1}{\sqrt{N}} 
\sum_{s,\vec{k}} \hat{\varepsilon}_{s,\vec{k}} e^{\imag \vec{k}\cdot \vec{R}_{i}} 
q_{s,\vec{k}}
\end{equation}
where $\hat{\varepsilon}_{s,\vec{k}}$ are the two-dimensional
polarization vectors (the electrons oscillate within the planes)
 and the vector $\vec{k}$ runs through the Brillouin zone
of the  \textit{three-dimensional} reciprocal lattice.  
This yields two branches $s=1,2$ 
of collective modes with eigenfrequencies 
$\omega_{s,\vec{k}}$, so that 
the vibrational energy per particle  can be expressed as: 
\begin{equation} \label{eq:E_V}
E_V =  \frac{1}{N} \sum_{s,\vec{k}} \frac{ \hbar \omega_{s,\vec{k}}}{2}
\end{equation}
It is useful to introduce the normalized density
of states (DOS) of the collective modes, that we write here in general
as: 
\begin{equation} 
\label{eqn:dos}
\rho(\omega) = \frac{1}{D N} \sum_{s=1}^D \sum_{\vec{k} \in BZ} \delta(\omega - \omega_{s,\vec{k}}) 
\end{equation}
($D$ is the number of branches, corresponding to the dimensionality
of the electron motion) as well its dimensionless moments:
\begin{equation} 
\label{eqn:Momenti}
M_n = \int \!\! d \omega \; \rho(\omega) \; { \left( \frac{\omega}{\omega_P}
  \right) }^n 
\end{equation}
with 
$\omega^2_P= 3 e^2 /( m r^{3}_{s} )$  
the usual 3D plasma frequency. 
With these definitions, the vibrational energy in
Eq.(\ref{eq:crysten}) 
is seen to be directly proportional to the first moment of the DOS,
with
\begin{equation}
  \label{eqn:E_GS_2}
  B(\gamma)= \frac{D \sqrt{3}}{2} M_1(\gamma)
\end{equation}
The usual 3D case in vacuum  is
recovered by restoring the
out-of-plane oscillations in Eq. (4), and by setting $D=3$ in
Eq. (\ref{eqn:E_GS_2}).
For example, for the BCC structure 
we find $M_1^{(3D)}=0.511$,  which yields the well
known   value  $B^{(3D)}=1.33$.
 \cite{Pollock_Hansen,Carr_Coldwell-Horsfall_Fein}

The analysis of the frequency spectrum shows that 
each given structure has a limited interval of mechanical
stability:  
for certain geometries, the dynamical
matrix acquires negative eigenvalues  around some critical
wavevector $k_c$, corresponding to purely imaginary
collective frequencies which drive the crystal unstable
(this phenomenon also exists  in free space, where FCC
and the simple cubic structure are known to be intrinsically unstable).
For example, in the interval of $\gamma$ under study, 
a structure with hexagonal symmetry is only
stable for $\gamma<1.32$, $3.5<\gamma<4.95$  and $5.05<\gamma<5.8$.

We have calculated the fluctuation term $B(\gamma)$ for the different 
symmetric structures (H, R, S, CR) identified in the previous section, 
 within their 
respective intervals of mechanical stability, as well as  for the
rhombic phase at $2.84<\gamma<2.86$ and $\gamma > 4.31$, which is shown in Fig. 
\ref{fig:B_coefficient}.a. 
As for the Madelung energy,  
two essentially different regimes can be identified.
For $\gamma\lesssim 1$, the electron motion is
mostly determined  by the Coulomb interactions within the
layers (interlayer forces are
negligible) and the collective modes of the pure 2D case are
recovered. If normalized by an appropriate  ``two-dimensional plasma
frequency'' $\omega_{2D}^2=e^2/mr_{s,2D}^3$, 
the first moment in the hexagonal phase
tends to the constant value $M_{1,2D}~=~0.814$.
\cite{Bonsall_Maradudin_Gann}
Going back to the present three-dimensional units, however,
where the moments are normalized as in Eq.(\ref{eqn:Momenti}),
the fluctuation term diverges  at
large separations as $B(\gamma)\simeq  \pi^{1/4} 3^{1/2} M_{1,2D}/2 \gamma^{1/2}$.
In the regime $\gamma\gtrsim 1$, on the other hand, 
the fluctuation term  flattens around a  value which roughly corresponds to
$2/3$ of the fluctuation 
in free space, indicated by the arrow in Fig. \ref{fig:B_coefficient}.a.
This follows from the fact that only the oscillations along 2
of the 3 space directions are allowed, as we can see explicitely from
Eq. (\ref{eqn:E_GS_2}).

\begin{figure}[htbp]
\hspace{-0.9cm}\includegraphics[scale=.37,angle=270.]{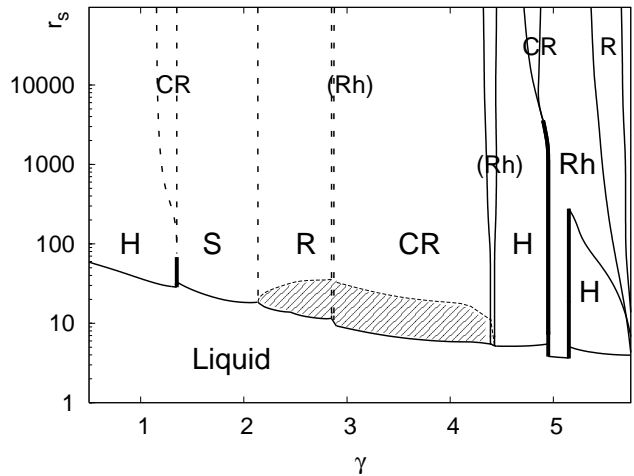}
\caption{Structural 
phase diagram  of the Wigner crystal in a layered environment, based on
the total energy (\ref{eq:crysten}), as a function of
  the anisotropy ratio $\gamma$ and the bulk density parameter $r_s$. 
The labels are the same as in previous figures.
The solid (dashed)  lines are for structural transitions where the crystal
parameters evolve  discontinuously (continuously).
The bold lines indicate mechanical instabilities, accompanied
by a discontinuity of the crystal energy.  
The melting line  is determined by solving Eq. (\ref{eq:rscrit}). For
the hatched region, see text.} 
\label{fig:evolution}
\end{figure}

The structural phase diagram resulting from the analysis of the total energy 
(\ref{eq:crysten}), including the vibrational term (\ref{eqn:E_GS_2}),
and taking into account 
the ranges of mechanical stability of the different phases, 
is reported in Fig. \ref{fig:evolution}. The first observation is that,
apart from the disappearance of the CR phase from certain intervals, 
which is penalized by its higher vibrational energy than the H phase,
the locus of the  structural transitions
does not change much with $r_s$. The sequence of phases identified in 
Fig. 1, based on the analysis of the Madelung energy, is recovered at
extremely large values of $r_s$. On the other hand,  the vibrational term 
affects the structural transitions already at $r_s\lesssim 1000$. 
This is due to the fact that,
even though the electrostatic term $A/r_s$ is still larger than the zero-point
fluctuation energy $B/r_s^{3/2}$, the latter undergoes much larger
relative variations among the different phases.
Below $r_s\sim 100$, 
the phase diagram 
is entirely determined by the
minimization of the vibrational energy (see
Fig. \ref{fig:B_coefficient}.a). As was stated above, however, the
overall shape of the phase diagram does not depend much on $r_s$,
the transitions being essentially determined by the parameter $\gamma$. 
Let us also  remark that the vibrational term is much less influenced
than the Madelung term  by the specific interlayer arrangements, 
whose effect (if any) is to slightly modify 
the range  of mechanical stability of each phase.

Another fundamental property of the system, which gives valuable
informations on the
collective vibrations of the particles, is 
the mean electronic fluctuation  $\left< u^2 \right>$. 
In the harmonic approximation, 
this quantity is proportional to the
inverse moment of the DOS of the collective modes, defined in
Eq. (\ref{eqn:Momenti}): 
\begin{equation} 
\label{eqn:u2}   
\left< u^2 \right> = 
\frac{1}{N}\sum_{k,s} \frac{1}{2\omega_{k,s}} =
\frac{D M_{-1}}{2\sqrt{3}}  r_s^{3/2}
\end{equation} 
where again we keep track of the explicit dependence on the
dimensionality $D$. As can be seen in  Fig.
\ref{fig:B_coefficient}.b,
it increases as each phase approaches  
the boundaries of its stability range. 
This is because the mechanical instabilities are  
approached  via a softening of a branch of collective modes, 
causing an increase of the DOS at low frequency
and, through Eq. (\ref{eqn:Momenti}), of the inverse moment $M_{-1}$.
A local increase also occurs  at the points where the 
staggered interlayer ordering is lost
(see e.g. the maximum at $\gamma=2.46$ within the R phase in
Fig. \ref{fig:B_coefficient}.b).

From analogous arguments, it follows from 
Eqs. (\ref{eqn:Momenti}) and  (\ref{eqn:E_GS_2}) 
that the vibrational energy  generally attains its minimum value 
close to  mechanical instabilities.
Within the present approximate framework, this
can cause the total energy to  jump discontinuously 
at the instability point when the next stable phase is attained, 
which corresponds to the bold lines is Fig. 3.
For example, the hexagonal lattice
becomes unstable at $\gamma>1.33$ and, for $r_s\lesssim 100$, the
transition to the square phase is accompanied by a small jump in
energy. 
Such discontinuities 
can in principle be avoided by allowing for Bravais lattices 
with  more than one electron per unit cell 
(the resulting internal structure 
could then be assimilated to some local tendency to electron pairing
\cite{Moulopoulos,Filippov}).
Note also that  it is precisely
close to mechanical instabilities, 
where $\langle u^2\rangle$ is largest,
that the neglected anharmonic corrections are expected to be most important. 
Their consequences on the structural phase diagram presented here
deserve further theoretical study.

\subsection{Melting of the crystallized state}

In this section, we  analyze the melting of the
crystallized state by making use of the  
Lindemann criterion, according to which
a transition to a liquid phase takes place when the  spread 
$\left< u^2 \right>$ 
attains some given fraction $\delta$ of the nearest-neighbor
distance $a_{n.n.}$. We  take  $\delta=0.28$ from Ref. \cite{Nagara},
which is appropriate for the quantum melting of both 2D and 3D
Wigner crystals. 
Solving the equation $\sqrt{\left< u^2 \right>}/a_{n.n.}=\delta$ in
terms of the density parameter $r_{s,2D}$ in the planes
leads to: 
\begin{equation}
\label{eq:rscrit}
r_{s,2D}^c = 
 \frac{M_{-1}(\gamma)}{
    2\delta^2 \mathcal{C}^2(\gamma)}  d^{1/2}
\end{equation}
where $\mathcal{C}=a_{n.n.}/r_{s,2D}$ is 
an aspect ratio relating the nearest-neighbor distance to 
the density parameter in the planes, and the implicit condition
$\gamma=\sqrt{\pi} r_{s,2D}^c/d$ holds. 
Note that for structures with rectangular symmetry, 
the Lindemann criterion must be modified
to account for the existence of two nonequivalent near-neighbor distances.
Here we use a simple 
generalization which consists in replacing $a_{n.n.}$ with the
average of the two shortest near-neighbor distances, and which reduces
to the ordinary criterion for the square and hexagonal structures. 
A check of the validity  of such generalized Lindemann
criterion will be given in Section II E, by direct
comparison with independent theoretical results on bilayer systems.

\begin{figure}[htbp]
\begin{flushleft}
\includegraphics[scale=.35,angle=270.]{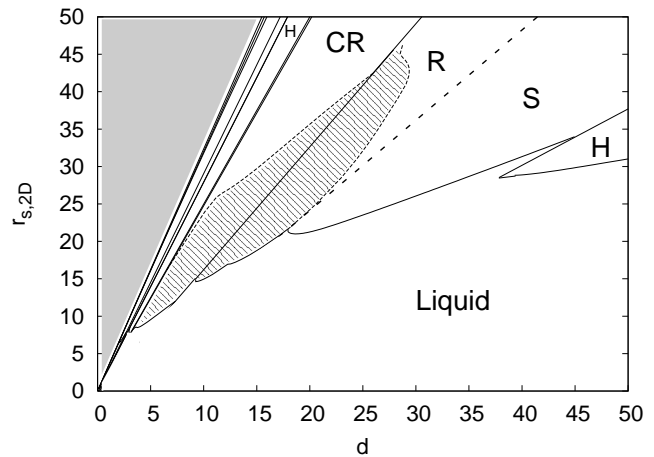} 
\caption{Phase diagram of the  Wigner crystal in a layered environment, 
as a function of the interlayer distance $d$ 
(in units of the effective Bohr radius $a_B^*$). 
The  discontinuity  close to the three-phase critical point 
(S, H, liquid) is due to the
different aspect ratios  $\mathcal{C}$ of the two competing structures.  
The hatched area is a region possibly characterized 
by an anisotropic liquid behavior (see text).
The shaded area corresponds to
$\gamma>6$ and  has not been studied. }  
\label{fig:melting}
\end{flushleft}
\end{figure}

The melting curve deduced from  Eq. (\ref{eq:rscrit}) 
for the different structures  considered here  
is illustrated in Figs.\ref{fig:evolution} and \ref{fig:melting}.
The most important result
is that the crystal melting  can be pushed to higher
densities by  reducing the interlayer spacing, which can already 
be inferred by neglecting the weak $\gamma$-dependence of the
coefficients $\mathcal{C}$ and $M_{-1}$ of Eq. (\ref{eq:rscrit}) in
the region $\gamma \gtrsim 1$.
The main reason to this is that for $\gamma \gtrsim 1$ the electron spread
is essentially governed by  \textit{three-dimensional} Coulomb
interactions, as we can see from the explicit dependence of
Eq. (\ref{eqn:u2}) on the bulk $r_s$, while the electron motion is
\textit{two-dimensional}, so that   
the appropriate nearest-neighbor distance 
for the Lindemann ratio is proportional to the
\textit{planar} density parameter $r_{s,2D}=
(2/\sqrt{3d})r_s^{3/2}$.\cite{ECRYS}

In addition, for each given spacing $d$, 
the geometrical confinement leads to a further 
stabilization of the crystal through a  
reduction of the spread $\langle u^2 \rangle$ itself. This effect is
directly reflected in Fig. \ref{fig:B_coefficient}.b in
a reduced value of $M_{-1}$ as compared to the corresponding value in
free space, and should not be confused with the trivial dimensional
factor $D$, that was taken out explicitely from Eq. (\ref{eqn:u2}).
It is due to the fact that, as soon as the cubic symmetry is lost, 
the restoring forces induced by the dipole-dipole interactions Eq. (5) are
not equivalent in the three space directions, so that  
the electron fluctuation becomes  anisotropic 
(the observed 
shrinking of the planar spread  would occur at the expense of  increasing 
the out-of-plane fluctuations, which are  anyhow 
suppressed in the model).
To give an example, taking an average value $M_{-1}\simeq 2.4$ 
and the aspect ratio  
$\mathcal{C}=\sqrt{\pi}$ for the square planar ordering 
yields a critical value  $r_{s,2D}^c\simeq 4.9 \sqrt{d}$.
Comparable results (within few percent) 
are found for the other structures.

In the opposite limit of  large separations ($\gamma\ll 1$), where
interlayer forces 
become negligible, we recover the usual critical value
$r_{s,2D}^c \simeq 40$ for the 2D hexagonal Wigner crystal.
Note that the actual critical value  at finite $\gamma$ always 
lies below this asymptotic estimate, 
confirming that the inclusion of interlayer interactions causes a 
stabilization of the crystal phase compared to the pure
two-dimensional case,
as was argued in the introduction.

\begin{figure}[htbp]
\begin{flushleft}
\centerline{\includegraphics[scale=.3,angle=270.]{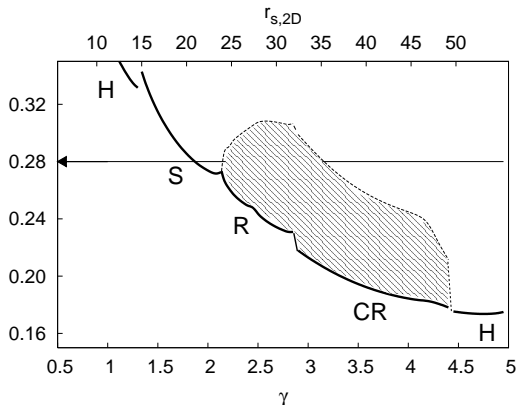}} 
\caption{Lindemann ratio $\sqrt{\langle u^2 \rangle}/a_{n.n.}$ 
 as a  function of $\gamma$ ($r_{s,2D}$, upper abscissa) 
at a given interlayer spacing $d=20$. The
continuous line is the average Lindemann ratio,  the
dashed line represents the Lindemann ratio along the direction of the 
closest near-neighbor (see text). The horizontal line sets the 
critical value for melting.
% (small discontinuities arise due to the
%different  aspect ratios  of the various structures H, S, R, CR).  
Upon increasing the electron density, 
the transition from the crystal to the liquid
could occur through an intermediate anisotropic liquid phase (hatched region,
see also Fig. 4).}
\label{fig:Lindemann}
\end{flushleft}
\end{figure}

A few comments on the limits of validity
of the present  model are in order.
First, 
the enhancement of Wigner crystallization predicted by Eq. (\ref{eq:rscrit})
cannot extend indefinitely: the melting line should 
eventually saturate at low separations when isotropic 
electron motion and three-dimensional screening are restored by
interlayer tunneling processes.\cite{contiIJMPB}    
On the other hand,
as was stated in the introduction, replacing
the host lattice of ions  by an effective jellium is allowed
provided that the spread of the electron wavefunction is larger than
the ion-ion distance $a_0$. From eq. (\ref{eqn:u2}), the condition
$\sqrt{\langle u^2\rangle} \gtrsim  a_0$  gives
$r_s\gtrsim 4$  ($r_{s,2D}\gtrsim 6$) for a typical value of $a_0=3
\AA$, assuming $\kappa=1$ and $m^*=m$. 
Below this value, the discrete nature of the host lattice should 
be included, which can further
stabilize the crystallized state, as pointed out in 
Refs.\cite{Pichard,Imada,Valenzuela}.

Before concluding this section, let us remark that, 
for anisotropic planar orderings such as
the rectangular and the centered rectangular structure, two
independent Lindemann ratios could in principle be defined (one for each
nonequivalent near-neighbor direction) rather than the single 
average criterion used so far.  
It would then appear that 
the melting along the short bonds is much easier than along the long
bonds, due to the closer overlap between the electron wavefunctions.
This phenomenon is illustrated in Fig. \ref{fig:Lindemann}, and could 
imply a tendency towards an anisotropic (or ``striped'') liquid phase,
 which is generally not ruled out by the 
isotropic nature of the Coulomb repulsion  (see also
the hatched regions in Figs. 3 and 4). 
\cite{Slutskin}
The results reported in Fig. 5 also indicate a possible reentrant behavior, 
although no conclusive answer can be given at this level of
approximation.

\subsection{Symmetric electron bilayer}
\label{sec:bilayer}

\begin{figure}[htbp]
\begin{flushleft}
\includegraphics[scale=.35,angle=270.]{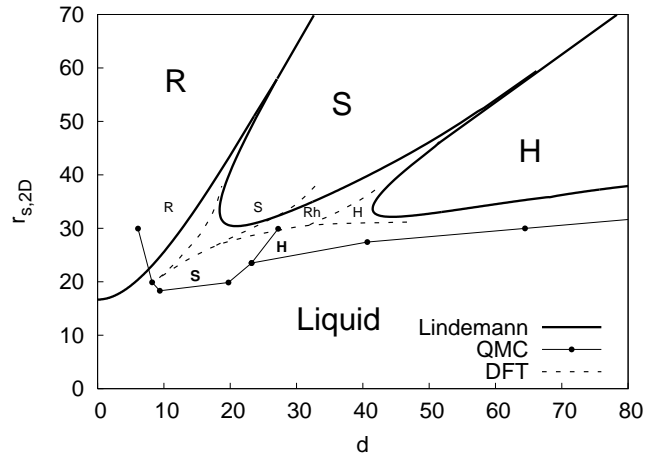}
\caption{
Phase diagram for the symmetric electron 
bilayer, in terms of the two dimensional
density parameter $r_{s,2D}$, as a function of the interlayer spacing $d$. 
The lengths are scaled to atomic units.
H,S,R denote respectively the 
Hexagonal, Square , Rectangular phase. 
Note that the QMC  simulation of
Ref. \cite{contiIJMPB}  was  restricted to study only two
phases (H,S), while an additional rhombic phase (Rh)
could be stabilized in the DFT calculations of
Ref.\cite{Goldoni_Peeters_2}.} 
\label{fig:bilayer}
\end{flushleft}
\end{figure}

In this section we analyze a system composed of two coupled electronic 
layers, in order to
check the validity of our approach by direct comparison with available
Density Functional Theory \cite{Goldoni_Peeters_2} and Quantum Monte Carlo
based calculations \cite{swierkowski,contiIJMPB}.
In the early work on classical bilayers, \cite{Goldoni_Peeters}
the analysis of the Madelung energy showed that
several structural phase transitions occur 
as the distance $d$ between the two planes is varied while keeping the 
electron density fixed. At short distances
the planar ordering is rectangular, and  collapses to the usual
hexagonal phase in the formal limit $d\to 0$. This phase evolves
continuously into a staggered square structure, when $d$ is of the order of the
interparticle spacing, which is clearly reminiscent of the BCC structure
observed in 3D space (cf. the discussion in Section II B). 
Upon further increasing $d$, the lattice
progressively deforms into a rhombic phase, to attain the hexagonal
staggered phase expected in the limit of 
independent layers.
Including the zero-point energy of the collective excitations as in
Eq. (\ref{eq:crysten}) raises the energy of the rhombic phase, which
therefore disappears from the phase diagram at sufficiently high
density, leaving the other transitions essentially unchanged.

We have analyzed the quantum melting of the different  Wigner crystal
structures realized in  such bilayer system by making
use of the Lindeman criterion discussed in the preceding Section.
We see from Fig. 6 that both the sequence of 
phases and the critical melting densities obtained within
the present quadratic approximation are in satisfactory agreement with the
more sophisticated numerical results of Refs. 
\cite{Goldoni_Peeters_2,swierkowski,contiIJMPB} 
(the  melting density is slightly underestimated as compared with QMC,
but quite similar to the DFT result).
It is interesting to see that the same trends observed in the 
preceding Section for the layered
solids are already present in the single bilayer.
In particular,  reducing the interlayer separation
leads to a sensible stabilization of the crystal  compared to the
isolated layers. This is clear in Fig. \ref{fig:bilayer}, where the
the melting line always lies below 
the critical value $r_{s,2D}^c\simeq
40$ of a purely 2D Wigner crystal. 
Note also that, contrary to Ref. \cite{swierkowski}, we find that 
the enhancement of Wigner crystallization is slightly more 
pronounced in an infinite array of layers than in a
single electron bilayer.

\section{Wigner crystallization in quasi one-dimensional solids}

We now extend our analysis to the case of quasi  one-dimensional
solids, which we model as periodic arrays of conducting wires. 
Following the general arguments presented in the previous Section, 
the enhancement of Wigner crystallization in this case 
should be even more pronounced than in
the two-dimensional case, because of the suppression of electronic
motion in two transverse directions rather than one.
The effect is even more dramatic if we consider that
a quantum crystal with genuine long-range order
cannot be realized in a pure one-dimensional system,\cite{Schulz} while it
is stabilized if we account for the  long-range Coulomb interactions 
between carriers on different wires.
\footnote{This can be understood by observing that the logarithmic 
divergence of
the mean spread $\langle u^2\rangle \sim \int d\omega\omega^{-1}$
arising in 1D is washed out in the present case where the wires are
embedded in a three-dimensional environment and momentum integrations run
over the 3D Brillouin zone.}

We shall  consider here a square array of wires for
illustrative purposes, although the 
specific arrangements occurring in real solids (rectangular,
rhombic) can be treated case by case.
Assuming a simple ordering of period $a$ within the wires and an
interwire distance $d$, 
the most general elementary three-dimensional Bravais lattice  
compatible with the given geometrical constraint
is described by the following basis vectors:
$\hat{A}_1 =(0,0,a)$, $\hat{A}_2 =(d,0,b)$,  $\hat{A}_3 =(0,d,c)$. 
The volume of the 3D unitary cell  is 
$V_c = a d^2  \equiv 4 \pi r^3_s/ 3$, 
the anisotropy ratio is now defined as $\gamma =  a / d$ and the 1D
density parameter is $r_{s,1D}=a/2$. 
As in the layered case, we take a compensating positive charge 
distributed uniformly in the bulk. 
The analysis presented 
in the preceding Section can be repeated here following the same steps:
i) calculation of the structure with the lowest 
Madelung energy upon varying the anisotropy ratio; 
ii) calculation of the corresponding vibrational energies;
iii) determination 
of the melting curve via the Lindemann criterion.
The generalization is straightforward, and we only report here the main 
results.

\begin{figure}[htbp]
\begin{flushleft}
\includegraphics[scale=.35,angle=270.]{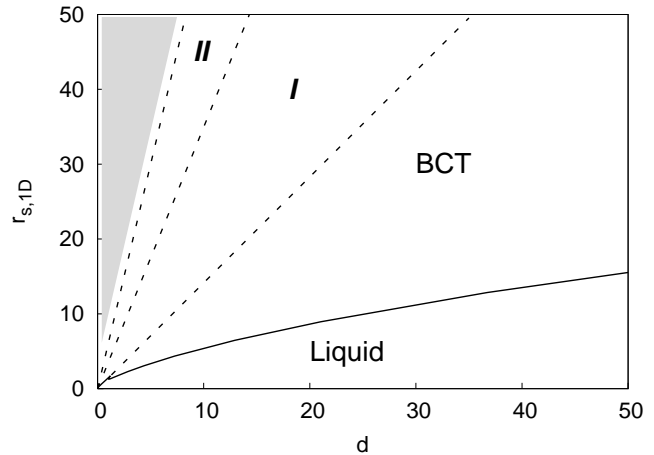}
\caption{Phase diagram   for a three-dimensional Wigner crystal
  embedded in a square array of 1-dimensional wires, of side  $d$.  
Lengths are scaled to effective atomic units.
For the definition of the phases I and II, see text. 
The shaded region corresponds to
$\gamma>10$ (not studied).}
\label{fig:chains}
\end{flushleft}
\end{figure}

The stuctural phase diagram (Fig.   \ref{fig:chains}) 
is clearly less rich than in the layered
case, because once the density and the interwire distance $d$ are fixed, only
the relative ordering between the electronic crystals on neighboring wires
remains to be determined, corresponding to  the pair of parameters $b$
and $c$.
In the limit $\gamma \rightarrow 0 $, the interwire interactions 
vanish and the limit of isolated wires is recovered:  
the Madelung constant $A$  diverges due to the 
isotropic distribution of the jellium, 
as explained previously (cf. footnote 43).  In this limit
the interwire ordering is staggered, with $b/a=c/a=1/2$, 
corresponding to a body centered
tetragonal (BCT) lattice
in three-dimensional space. 
The BCT structure, everywhere compatible with 
a square array of wires,  has the lowest Madelung energy 
in the whole range $0<\gamma<2.83$, with the
two special values  $\gamma^*=\sqrt{2}$  and  $\gamma^*=2$ corresponding
respectively to a BCC and a FCC.
For $\gamma>2.83$ the minimum configuration becomes less symmetric,
with $c/a\neq 1/2$ but the ratio $b/a$ still locked to the value $1/2$
up to $\gamma=6.99$. 
This phase is denoted (I) in Fig.\ref{fig:chains}.  
Beyond $\gamma=6.99$, a second structural transition occurs leading to a
generic phase (II) 
with both $b/a\neq 1/2$ and $c/a \neq 1/2$. Other transitions can take place 
at larger values of $\gamma$, within the generic phase II.
The sequence of phases does not change upon inclusion of the
vibrational term.

By applying the Lindemann rule we obtain a parametric 
formula for the melting curve analogous to Eq. (\ref{eq:rscrit}):
\begin{equation}
r_{s,1D}^c = \frac{1}{(128 \pi)^{1/3}} 
\left[\frac{M_{-1}(\gamma)}{\delta^2}\right]^{2/3} \; d^{2/3}
\label{eq:Lind1D}
\end{equation}
with the implicit condition 
$\gamma=r_{s,1D}^c/2d$.
%$\gamma = (4 \pi /3)  {\left( r_s / d \right)}^{3}$.
The consequences of geometrical confinement 
evidenced in the layered case are recovered here.  
The electron spread along the wires is again governed by the
three-dimensional plasma frequency [cf. Eq. (\ref{eqn:u2})], 
due to  the isotropic nature of
the Coulomb interactions, while the nearest-neighbor distance here scales
with $r_{s,1D}\equiv (2\pi/3)r_s^3/d^2$. 
Further stabilization of the crystallized state is achieved through 
a reduction of the electron spread along the wires, revealed by 
an 
inverse moment $M_{-1}$ which is typically $50\%$ lower than the value
in vacuum.
Its $\gamma$-dependence for $\gamma\gtrsim 1$  
is quite flat  (not shown), except in the vicinity of the transition at
$\gamma=2.85$, where it raises due to the mode softening 
discussed in Section II C.
Replacing the average value 
$M_{-1}\simeq 2$  into Eq. (\ref{eq:Lind1D}) yields 
$r_{s,1D}^c\simeq 1.2 d^{2/3}$,
corresponding to an even stronger
enhancement of Wigner crystallization 
than in the layered case (see  Table \ref{tab:1}).  
In the opposite anisotropic limit $\gamma \ll 1$, $M_{-1}$
diverges as in the case of an isolated wire
%for the reasons explained above 
(cf. footnote 43),  so
that the Wigner crystal is never stabilized ($r_{s,1D}^c \to \infty$).

\begin{table}[htbp]
  \centering
  \begin{tabular}{c|c|c|c|c} 
  \hline \hline
  &  $\gamma$   & crystal melting & $d=8 a_B^*$ & $d=20 a_B^*$\\ \hline  
  layers & $\sqrt{\pi} r_{s,2D}/d$ 
         & $ r_{s,2D}^c \simeq 4.9 \ d^{1/2}$ \hfill  $\gamma\gtrsim 1$ &
           $14$ & $ 21$\\
  & &  \hfill  $\: \: \: \simeq 40 $ \hfill  $\: \: \gamma\ll 1$ & & \\ \hline
  wires & $2 r_{s,1D}/d$
         & $r_{s,1D}^c\simeq 1.2 \ d^{2/3}$ \hfill $\gamma\gtrsim 1$ &
         $ 5$ & $9$\\
  & & \hfill $\: \: \: \to \infty $ \hfill $ \: \: \gamma \ll 1$ & &
  \\\hline \hline 
  \end{tabular}
  \caption{Definition of the anisotropy ratio $\gamma$, approximate melting
    lines obtained for quasi two-dimensional and quasi one-dimensional
    systems, and specific values obtained at two different interlayer   
    (interwire) distances $d$, expressed  in units of 
    the effective Bohr radius $a_B^*$ 
    (right columns). } 
\label{tab:1}
\end{table}

\section{Conclusions}

We have investigated the Wigner crystallization of electrons
in quasi low-dimensional compounds, where the carrier motion is effectively
low-dimensional, while the Coulomb interactions are
assumed long-ranged and isotropic. The system properties are found to
depend crucially on the ratio $\gamma$ 
of the mean interparticle spacing within the conducting units (layers or
chains) to the separation $d$ between units.
%to the relative separation 
%$d$ between conducting units (layers or chains).
While the behavior expected for isolated units is 
recovered at large separations ($\gamma \ll 1$),
an overall isotropic  ordering of the charges is achieved for
$\gamma\gtrsim 1$,   when the 
interactions between different units become important.
%distance $d$ is comparable or 
%less than the average interparticle spacing. 
In this case, three-dimensional structures 
as close as possible to the ideal case of a BCC are formed, leading 
to a cascade of structural transitions 
which can be tuned by varying the particle density, or the distance
$d$ itself. In addition to this rich phase diagram, 
the presence of isotropic Coulomb interactions in such anisotropic
compounds results in 
a strong stabilization of the charge ordered phases, 
possibly up to densities of practical interest, where the characteristic 
energy scales of the Wigner crystal can become comparable with other 
relevant scales in the solid.
Although it is clear that the interplay with several other
factors such as the periodic lattice
potential,\cite{Tsiper,Imada,Baer05,Pichard,Hubbard78,Valenzuela} 
chemical impurities,\cite{Chui}  polarons\cite{holecrystal,PWC} or 
magnetic interactions\cite{cuprates,Castroneto} 
should be considered for an accurate description of real materials, 
 the long-range Coulomb
interactions appear in light of the present study 
as a key ingredient  to understand the charge ordering
phenomena in quasi low-dimensional systems.

\vspace{.2cm}

\section*{ACKNOWLEDGMENTS}

We thank S. Ciuchi for critical and constructive discussions. 
G.\ R.\ thanks the kind hospitality of CNRS-LEPES Grenoble 
(France) and finantial support 
by MIUR-Cofin 2004/2005
matching funds programs.


\begin{thebibliography}{99}

\bibitem{Wigner} E.\ P.\ Wigner, Phys.\ Rev.\ {\bf 46}, 1002 (1934), 
E.\ P.\ Wigner, Trans.\ Faraday.\  Soc.\ {\bf 34}, 678 (1938).



\bibitem{Crandall_Williams} R.\ S.\ Crandall, R.\ Williams, Phys.\
  Lett.\ A {\bf 34} 404 (1971).

\bibitem{Grimes_Adams} C.\ C.\ Grimes, G.\ Adams, Phys.\ Rev.\ Lett.\
  {\bf 42}, 795 (1979).

\bibitem{Andrei_book} E.\ Y.\ Andrei,  {\em  Two-Dimensional Electron
    Systems on Helium and other Cryogenic Substrates}, (Kluwer 
Academic Publ. 1997), pags. 245-279 and refs. therein.


\bibitem{Yoon} J.\ Yoon, C.\ C.\ Li, D.\ Shahar, D.\ C.\ Tsui,
  M.\ Shayegan, Phys.\ Rev.\ Lett. {\bf 82}, 1744 (1999).

\bibitem{Hubbard78}  J. Hubbard, Phys. Rev. B 17, 494 (1978).


\bibitem{Monceau} F. Nad, P. Monceau, C. Carcel, and
  J. M. Fabre, Phys. Rev. B 62, 1753 (2000); 
P. Monceau, F. Ya. Nad and S. Brazovskii,
  Phys. Rev. Lett. 86, 4080 (2001)

\bibitem{Chow} D. S. Chow et al., Phys. Rev. Lett. 85, 1698
  (2000) 


\bibitem{Hiraki-Nakazawa} K. Hiraki and K. Kanoda,
  Phys. Rev. Lett. 80, 4737 (1998); Y. Nakazawa et al.,
  Phys. Rev. Lett. 88, 076402 (2002)

\bibitem{Abbamonte} P. Abbamonte et al., Nature 431, 1081 (2004)
\bibitem{Abbamonte2} A. Rusydi et al., cond-mat/0511524 

\bibitem{Horsch}P. Horsch, M. Sofin, M. Mayr, and M. Jansen,
  Phys. Rev. Lett. 94, 076403 (2005)

\bibitem{Drichko} M. Dressel and  N. Drichko, Chem. Rev. 104, 5689 (2004)

\bibitem{cuprates}  H. C. Fu, J. C.  Davis, D.-H. Lee,
  cond-mat/0403001 (unpublished) 


\bibitem{holecrystal} G.\ Rastelli, S.\ Fratini, P.\
  Qu\'emerais, Eur.\ Phys.\ J.\ B {\bf 42}, 305 (2004)

\bibitem{Baer05} D. Baeriswyl and S. Fratini, J. Phys IV France 131,
  247 (2005) 
\bibitem{Pichard} H. Falakshahi et al.,  Eur. Phys. J. B 39, 93 (2004)

\bibitem{Valenzuela} B. Valenzuela, S. Fratini and D. Baeriswyl, 
Phys. Rev. B 68, 045112 (2003) 




\bibitem{Goldoni_Peeters} G.\ Goldoni, F.\ M.\ Peeters,  Phys.\ Rev.\
  B {\bf 53}, 4591 (1996).

\bibitem{Schweigert_Schweigert_Peeters} I. V. Schweigert,
  V. A. Schweigert, F. M. Peeters, Phys.  Rev.  Lett { \bf 82}, 5293
  (1999); 
I. V. Schweigert, V. A.
  Schweigert, F. M. Peeters, Phys.\ Rev.\ B {\bf 60}, 14665 (1999) 



\bibitem{Goldoni_Peeters_2} G.\ Goldoni, F.\ M.\ Peeters,  Europhys.\ Lett.\  {\bf 37}, 293 (1997).%


\bibitem{swierkowski} L. \'Swierkowski, D. Neilson and
  J. Szyma\'nski, Phys. Rev. Lett. 67, 240 (1991)

\bibitem{contiIJMPB} G. Senatore, F. Rapisarda and S. Conti,
  Int. J. Mod. Phys. B 13, 5-6, 479 (1999)





\bibitem{Carr} W.\ J.\ Carr, Phys.\ Rev.\ {\bf 122}, 1437 (1961).

\bibitem{Carr_Coldwell-Horsfall_Fein} W.\ J.\ Carr, R.\ A.\
  Coldwell-Horsfall, and A.\ E.\ Fein, Phys.\ Rev.\ {\bf 124}, 747
  (1961).
  
\bibitem{Clark} C.\ B.\  Clark, Phys.\ Rev.\ {\bf 109}, 1133 (1958).


\bibitem{PWC}  S. Fratini, P. Qu\'emerais, Eur. Phys. J. B 14, 99
  (2000); \textit{idem}, Eur. Phys. J. B 29, 41 (2002); G. Rastelli,
  S. Ciuchi, Phys. Rev. B 71, 184303 (2005)

\bibitem{Ewald} P.\ P.\ Ewald,  Ann.\ Phys.\ {\bf 64}, 253 (1921).

\bibitem{Sherman} E. Ya. Sherman, Phys. Rev. B 52, 1512 (1995)
\bibitem{Tsiper} E. V. Tsiper and A. L. Efros, Phys. Rev. B 57, 6949 (1998)

\bibitem{Pollock_Hansen}  E.\ L.\ Pollock, J.\ P.\ Hansen, Phys.\ Rev.\ A {\bf 8}, 3110 (1973);
R.\ Mochkovitch, J.\ P.\ Hansen, Phys.\ Letts.\ A {\bf 73}, 35 (1979).

\bibitem{Bonsall_Maradudin_Gann} L.\ Bonsall, A.\ A.\ Maradudin, Phys.\ Rev.\ B {\bf 15}, 1959 (1977); 
R.\ C.\ Gann, S. Chakravarty  G.\ V.\ Chester,  Phys.\ Rev.\ B {\bf 20}, 326 (1979). 
\bibitem{Moulopoulos} K. Moulopoulos, N. W. Ashcroft, Phys. Rev. B 48,
  11646 (1993)

\bibitem{Filippov} Yu. G. Pashkevich and A. E. Filippov, Phys. Rev. B
  63, 113106 (2001) 

\bibitem{Nagara} H. Nagara, Y. Nagata and T. Nakamura, 
Phys.\ Rev.\ A {\bf 36}, 1859 (1987).

\bibitem{ECRYS} G. Rastelli, P. Qu\'emerais and S. Fratini,
  J. Phys. IV France 131, 277 (2005)







\bibitem{Slutskin} A.\ A.\ Slutskin, V.\ V.\ Slavin, and H.\ A.\ Kovtun, Phys.\ Rev.\ B {\bf 61}, 14184 (2000)

\bibitem{Imada} Y.\ Noda, M.\ Imada, Phys.\ Rev.\ Lett. {\bf 89}, 176803 (2002)

\bibitem{Schulz} H. J. Schulz, Phys. Rev. Lett. 71, 1864 (1993)

\bibitem{Chui} S. T. Chui and B. Tanatar, Phys. Rev. Lett. 74, 458 (1995)

\bibitem{Castroneto} B. P. Stojkovic, Z. G. Yu, A. R. Bishop,
  A. H. Castro Neto and N. Grnbech-Jensen, Phys. Rev. Lett. 82, 4679 (1999); 
V. M. Pereira, J. M. B. Lopes dos Santos, A. H. Castro Neto, cond-mat/0505741




\end{thebibliography}
\end{document}